\documentclass{elsart}
\usepackage{graphicx}
\usepackage{amsmath}
\usepackage{amssymb}
\usepackage{epsfig}
\usepackage{dcolumn}
\usepackage{bm}
\begin{document}
\begin{frontmatter}

\title{Complexity of Self-similar Hierarchical Ensembles}
\author{A.I. Olemskoi}
\address{Applied Physics Institute, Petropavlovskaya St.,  58, Sumy, 40030,
Ukraine}
\ead{alex@ufn.ru}
\date{}

\begin{abstract}
Within the framework of generalized combinatorial approach, the complexity is
determined for infinite set of self-similar hierarchical ensembles. This
complexity is shown to increase with strengthening of the hierarchy coupling to
the value, which decreases with growth of both scattering of this coupling and
non-extensivity parameter.
\end{abstract}

\begin{keyword}
Self-similar hierarchical ensemble; complexity; $q$-multinomial coefficient.
 \PACS 02.50.-r, 05.20.Gg, 05.40.-a.
\end{keyword}
\end{frontmatter}

Hierarchical structure is one of the most universal peculiarities of complex
systems in physics, biology, economics and so on \cite{1}--\cite{5}. Their
evolution is shown \cite{6} to reduce to anomalous diffusion process in
ultrametric space of the self-similar hierarchical system, whose steady-state
distribution over hierarchical levels is given by the Tsallis power-law, being
inherent in non-extensive systems \cite{7}. Well-known feature of hierarchical
systems consists in that every of statistical ensembles of given level breaks
with passage into lower one to subensembles, then every of these breaks to more
small subensembles of the following level, and so on (see, for example, Figure
1).
\begin{figure}[htb]
\centering
\includegraphics[width=130mm,]{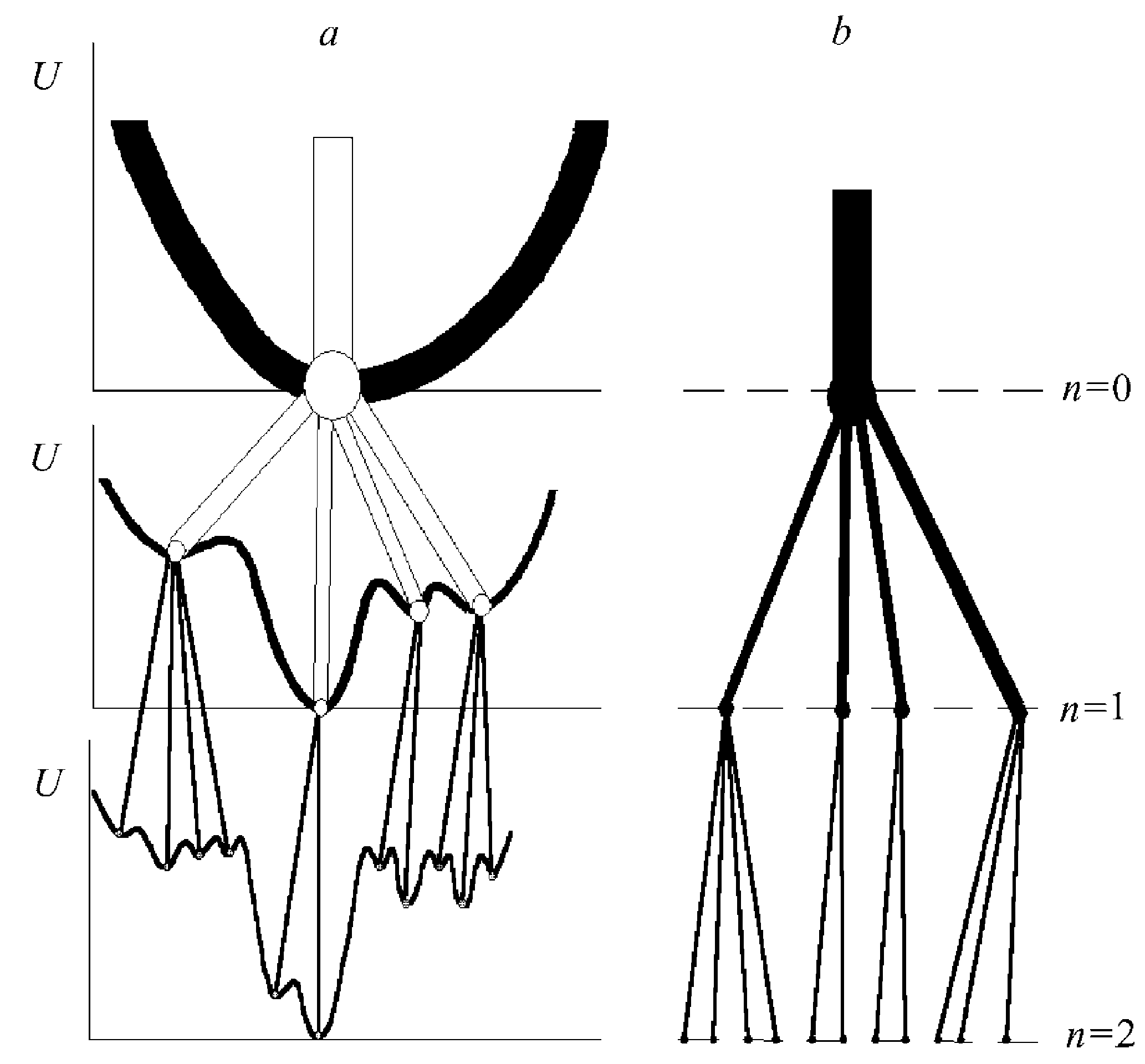}
\caption{Characteristic form of the internal energy relief (a) and related
hierarchical tree (b) of complex system \cite{3}}
\end{figure}
From the statistical point of view, the set of above (sub)ensembles is
characterized by the {\it complexity}, whose value determines the scattering of
the hierarchical coupling -- in analogy with the entropy in usual statistical
systems. This article aims to determine the complexity for self-similar
hierarchical ensembles.

According to Ref.\cite{7} the non-extensive statistics is based on the
definitions of both logarithmic and exponential functions by the equations
\begin{equation}
\ln_q(x):= \frac{x^{1-q}-1}{1-q},\ \exp_q(x):= \left[1+(1-q)x\right]_+^{1\over
1-q};\quad [y]_+:= \max(0,y),\ q\leq 1,
 \label{1}
\end{equation}
that are reduced to the usual functions in the limit $q\to 1$. Introducing
$q$-deformed production and ratio of positive values $x, y$ as follows:
\begin{equation}
x\otimes_q y=\left[x^{1-q}+y^{1-q}-1\right]_+^{1\over 1-q},\ x\oslash_q
y=\left[x^{1-q}-y^{1-q}+1\right]_+^{1\over 1-q};\quad x,y>0,
 \label{2}
\end{equation}
it is easily to convince that these satisfy to usual properties
$\ln_q(x\otimes_q x)=\ln_q x+\ln_q y$, $\ln_q(x\oslash_q x)=\ln_q x-\ln_q y$;
$\exp_q(x)\otimes_q\exp_q(y)=\exp_q(x+y)$,
$\exp_q(x)\oslash_q\exp_q(y)=\exp_q(x-y)$.

Within the framework of combinatorial approach \cite{8}, the $q$-deformed
statistics is reduced to consideration of the generalized factorial
$N!_q:=1\otimes_q\dots\otimes_q N$, whose logarithm is as follows:
\begin{equation}
\ln_q(N!_q)=\frac{\sum_{i=1}^{N}i^{1-q}-N}{1-q}.
 \label{3}
\end{equation}
In thermodynamic limit $N\to\infty$, above sum is estimated by related
integral, whose calculation gives
\begin{eqnarray}
\ln_q(N!_q)=\left\{
\begin{array}{ll}\frac{N}{2-q}\ln_q N-\frac{N}{2-q}+{\it O}(\ln_q N),\qquad q\ne 2,\\
N-\ln N+{\it O}(1),\quad\quad\quad\quad\quad\quad q=2.
\end{array} \right.
\label{4}
\end{eqnarray}
Defining $q$-deformed multinomial coefficient by the equation
\begin{equation}
{N\choose N_1\dots N_k}_q
:=(N!_q)\oslash_q\left[(N_1!_q)\otimes_q\dots\otimes_q(N_k!_q)\right],
 \label{5}
\end{equation}
where the set of integers $N_i$ satisfies to the condition
$N=\sum_{i=1}^{n}N_i$, we find
\begin{equation}
{N\choose N_1\dots N_k}_q
=\left[\sum\limits_{i=1}^{N}i^{1-q}-\sum\limits_{i_1=1}^{N_1}i_1^{1-q}-\dots
-\sum\limits_{i_k=1}^{N_k}i_k^{1-q}+1\right]_+^{1/(1-q)}.
 \label{6}
\end{equation}
From here, similarly to Eq.(\ref{4}), one obtains the expression
\begin{eqnarray}
\ln_q{N\choose N_1\dots N_k}_q\simeq\left\{
\begin{array}{ll}\frac{N^{2-q}}{2-q}C_{2-q}\left(\frac{N_1}{N},\dots,\frac{N_k}{N}\right),\qquad q>0,q\ne 2,\\
-C_1(N)+\sum\limits_{i=1}^{k}C_{1}(N_i),\quad\quad\quad\quad\quad\quad q=2
\end{array} \right.
\label{7}
\end{eqnarray}
for the Tsallis entropy
\begin{equation}
C_q(p_1,\dots,p_N):=-\sum\limits_{i=1}^{N}p_i\ln_q
p_i=-\frac{\sum_{i=1}^{N}p_i^q-1}{1-q}.
 \label{8}
\end{equation}

Above formalism is generalized easily for consideration of the hierarchical
systems \cite{9}. Assume that the $N$ states of upper level are distributed
over the ensembles $i=1,\dots,n$, every of which contains $N_i$ states. In
turn, $N_i$ states are bunched in $m_i$ subensembles $ij$, every of which
contains $N_{ij}$ states, where the relations $\sum_{j=1}^{m_i}N_{ij}=N_i$,
$\sum_{i=1}^{n}N_i=N$ are fulfilled. Then, instead of the multinomial
coefficient (\ref{5}) it is necessary to use the expression
\begin{equation}
{N\choose N_{11}\dots N_{nm_n}}_q={N\choose N_{1}\dots
N_n}_q\otimes_q{N_1\choose N_{11}\dots
N_{1m_1}}_q\otimes_q\dots\otimes_q{N_n\choose N_{n1}\dots N_{nm_n}}_q,
 \label{9}
\end{equation}
whose $q$-logarithm is
\begin{equation}
\ln_q{N\choose N_{11}\dots N_{nm_n}}_q=\ln_q{N\choose N_{1}\dots
N_n}_q+\sum\limits_{i=1}^{n}\ln_q{N_i\choose N_{i1}\dots N_{im_i}}_q.
 \label{10}
\end{equation}
As a result, using estimation (\ref{7}) arrives at the connection between the
complexities of the nearest hierarchical levels
\begin{equation}
C_Q\left(p_{11},\dots,p_{nm_n}\right)=C_Q\left(p_{1},\dots,p_{n}\right)
+\sum\limits_{i=1}^{n}p_{i}^Q
C_Q\left(\frac{p_{i1}}{p_i},\dots,\frac{p_{im_i}}{p_i}\right),
 \label{11}
\end{equation}
whose distributions over states are given by the equations $p_{ij}=N_{ij}/N$,
$p_i=N_{i}/N$ (here we introduce 'physical' non-extensivity parameter $Q=2-q$,
$1\leq Q\leq 2$). Using definition (\ref{8}) and connections
$p_i=\sum_{j_i=1}^{m_i}p_{ij_i}$ at condition $p_i-p_{ij_i}\ll p_i$ gives
estimation
\begin{equation}
C_Q\left(\frac{p_{i1}}{p_i},\dots,\frac{p_{im_i}}{p_i}\right) \approx
\frac{Q}{2}\sum_{j_i=1}^{m_i}\left(\frac{p_i-p_{ij_i}}{p_i}\right)^2,
 \label{12}
\end{equation}
with whose accounting one finds
\begin{equation}
C_Q\left(p_{11},\dots,p_{nm_n}\right)-C_Q\left(p_{1},\dots,p_{n}\right)\approx
\frac{Q}{2}\sum_{i=1}^{n}p_i^{Q-2}\sum_{j_i=1}^{m_i}\left(p_i-p_{ij_i}\right)^2.
 \label{13}
\end{equation}
If statistical states are distributed within microcanonical ensembles, then
both probabilities and related complexities are determined by the level number
$n$: $\{p_{i}\}_{1}^{n}\Rightarrow p_{n}$, $\{p_{ij_i}\}_{1}^{m_i}\Rightarrow
p_{n+1}$; $C_Q\left(p_{1},\dots,p_{n}\right)\Rightarrow C(n)$,
$C_Q\left(p_{11},\dots,p_{nm_n}\right)\Rightarrow C(n+1)$. As a result,
Eq.(\ref{13}) takes the simplest form:
\begin{equation}
C(n+1)-C(n)\approx \frac{Q(n+1)}{2}p_n^{Q-2}\left(p_n-p_{n+1}\right)^2.
 \label{14}
\end{equation}

At fixed level number $n\gg 1$, relation obtained is reduced to the
differential equation
\begin{equation}
\frac{\partial^2 C(n)}{\partial p^2_n}= Q(n+1)~p_n^{-(2-Q)},
 \label{15}
\end{equation}
whose integration gives the dependence
\begin{equation}
C(n)=p_0^Q-\frac{Q}{Q-1}~p_0^{Q-1}p_n+\frac{n+1}{Q-1}~p_n^Q
 \label{16}
\end{equation}
at the boundary conditions
\begin{equation}
C(n=0)=0,\quad \left.\frac{\partial C(n)}{\partial p_n}\right|_{n=0}=0.
 \label{17}
\end{equation}
Then, using self-similar distribution \cite{6}
\begin{equation}
p_n=A\left[\Delta+(Q-1)(n+1) \right]^{-{1\over Q-1}},\quad A\equiv
(2-Q)\left[(Q-1)+\Delta\right]^{\frac{2-Q}{Q-1}},
 \label{18}
\end{equation}
where parameter $\Delta$ determines the scattering over hierarchical levels, we
obtain the complexity distribution shown in Figure 2. \begin{figure}[htb]
\centering
\includegraphics[width=100mm]{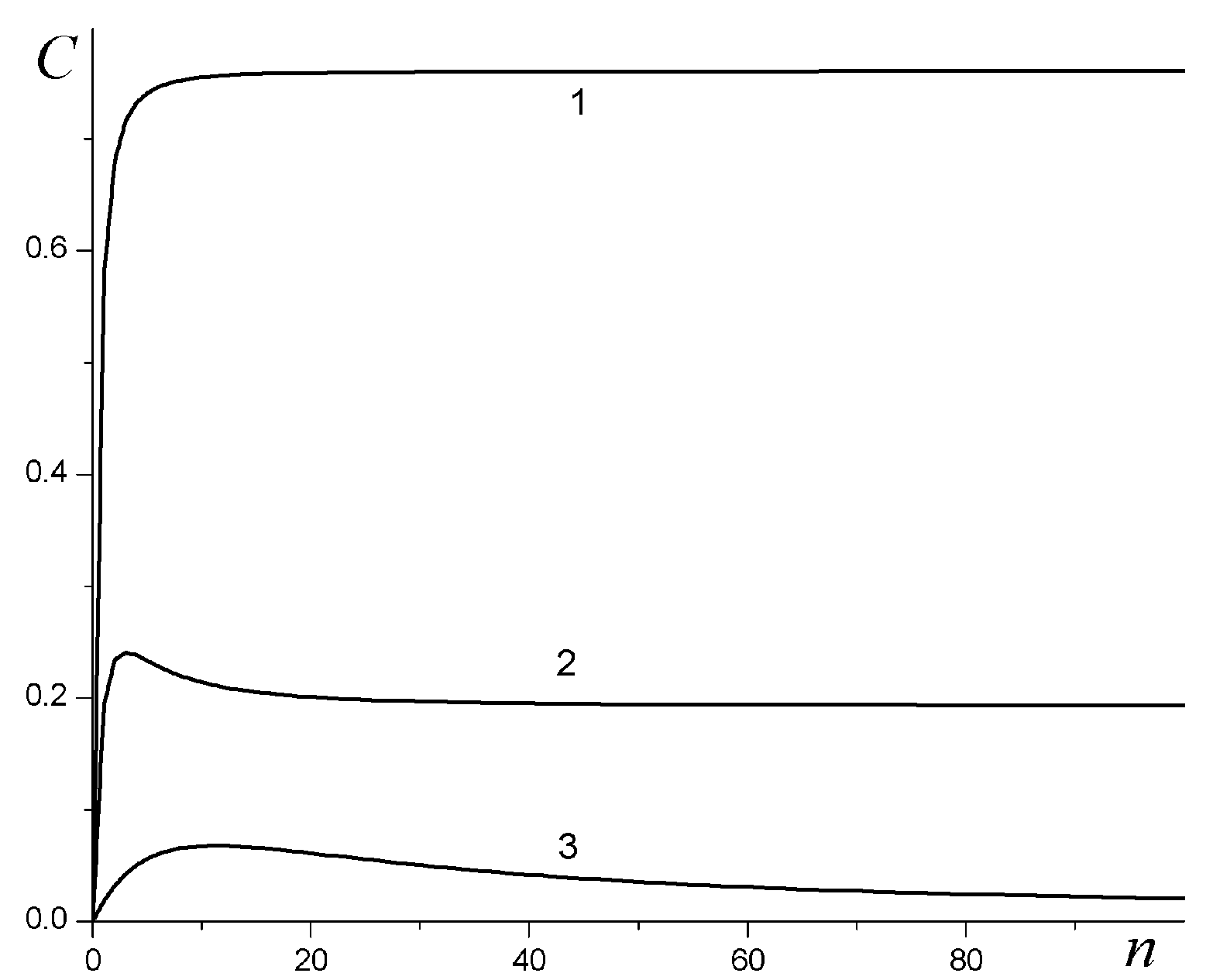}
\caption{Complexity dependence on hierarchy level number (curves 1, 2, 3 relate
to parameters $\Delta=0.1,\ 1.0,\ 10$; $Q=1.5$)}
\end{figure}
As it should be, on the upper level $n=0$, where only one statistical ensemble
takes place, the complexity is $C=0$. With passage to lower levels the value
$C$ increases, taking maximum value at $n_0=(Q-1)+\Delta$, after which
characteristic magnitude $C_\infty=p_0^Q$ is reached. However, it follows to
take into account that above consideration is applicable for level numbers
$n\gg 1$ only. Therefore, the pointed out maximum is displayed at condition
$\Delta\gg 1$, when distribution (\ref{18}) differs very weakly from the
exponential one, being inherent in usual additive systems. At moderate
scattering of the hierarchical coupling $\Delta\sim 1$, the system complexity
increases fast from zero to the limit value
\begin{equation}
C_\infty=\left[\frac{2-Q}{(Q-1)+\Delta}\right]^Q,
 \label{19}
\end{equation}
which decreases with growth of both scattering $\Delta$ and non-extensivity
parameter $1\leq Q\leq 2$.

Evolution process of the hierarchical ensemble is determined by both diffusion
time $\tau_d=(\Delta^{2-Q}/D)n^Q\tau_0$ and scale $\tau=n^2\tau_0$ of the
steady-state reaching, where $D$ is diffusion coefficient, $\tau_0$ is time of
passage between nearest levels \cite{6}. At $t\ll\tau_d$, the diffusion over
the hierarchical tree has anomalous character to be determined by the law
$n(t)\approx Q^{1/Q}(t/\tau_0)^{1/Q}$. With time growth within the interval
$\tau_d\sim t\ll\tau$, when contributions of anomalous drift and diffusion are
comparable, the passage into the normal regime $n(t)\approx \sqrt{2(t/\tau_0)}$
happens. At $t\ll\tau$, the distribution over hierarchical levels tends to the
steady-state law (\ref{18}) according to the time dependence
\begin{equation}
p_n(t)=p_n\left[1-\left(t/\tau_n\right)^{-\frac{1}{Q-1}}\right],
 \label{20}
\end{equation}
which is reduced to the value $p_n(t)=0$ at $t\leq\tau_n$,
$\tau_n\equiv(\Delta/QD)n\tau_0$, $n\ne 0$ (here, initial distribution
$p_n(t=0)=\delta_{n0}$ is taken in form of the Kronecker $\delta$-symbol).

Non-stationary complexity is determined by Eq.(\ref{15}), where one should take
the boundary conditions
\begin{equation}
C(n,t=0)=0, \quad C(n,t=\infty)=C(n),\qquad n\ne 0
 \label{21}
\end{equation}
instead of Eqs.(\ref{17}). As a result, we derive the time-dependent complexity
in the form
\begin{equation}
C(n,t)=\left(\frac{p_0^Q}{p_n}-\frac{Q}{Q-1}~p_0^{Q-1}\right)p_n(t)
+\frac{n+1}{Q-1}~p_n^Q(t),\quad n\gg 1,
 \label{22}
\end{equation}
being generalization of the steady-state distribution (\ref{16}). Inserting
Eqs.(\ref{18}), (\ref{20}) into Eq.(\ref{22}), we obtain the time dependence of
the complexity shown in Figure 3. During the ballistic interval $t\leq\tau_n$,
the complexity keeps the initial value $C=0$, whereas in the course of the time
$t>\tau_n$ it increases fast before the steady-state value (\ref{16}), being
bounded by the maximum value (\ref{19}).
\begin{figure}[htb]
\centering
\includegraphics[width=100mm]{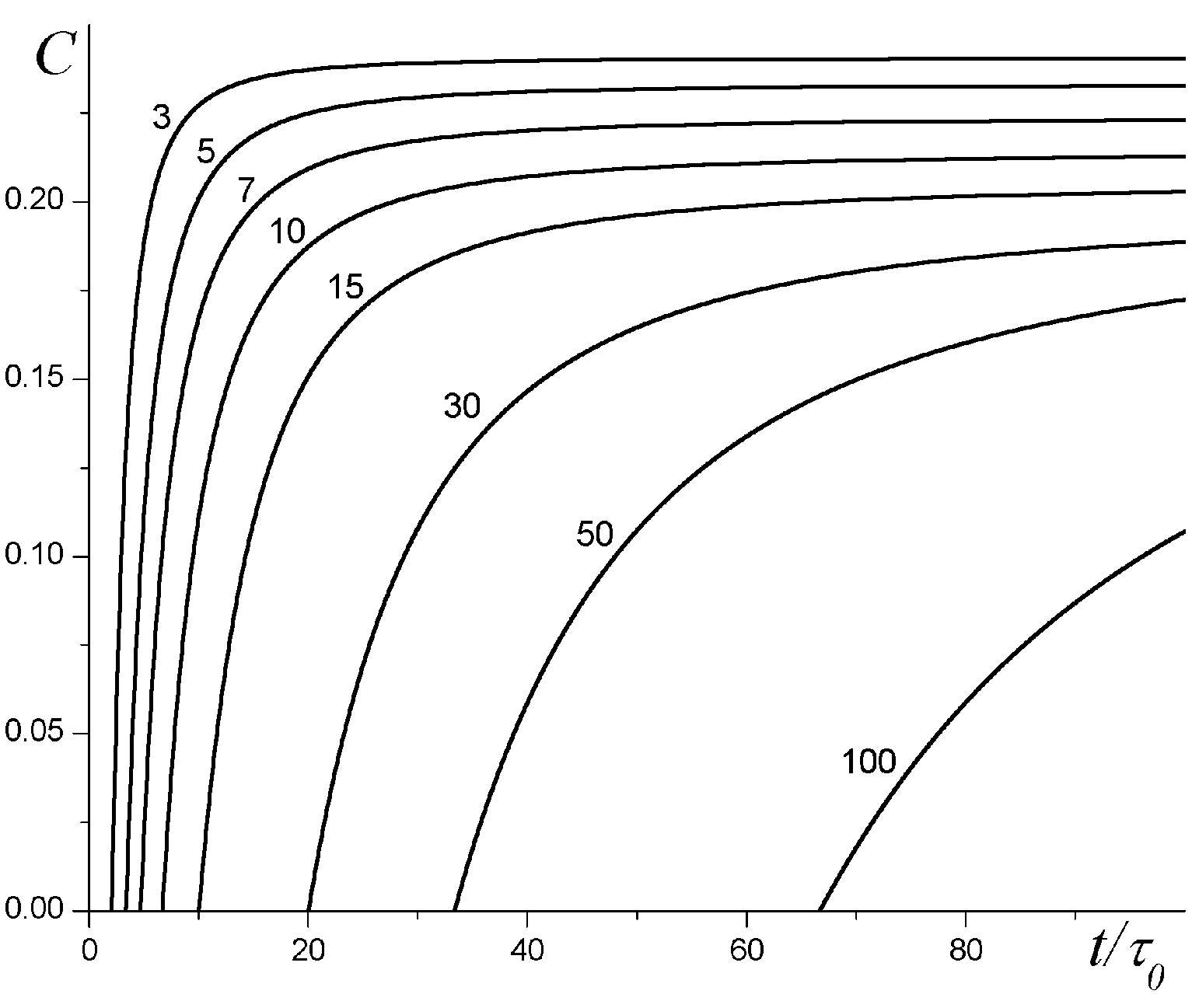}
\caption{Time dependencies of the complexity on different hierarchy levels
(their numbers are noticed near curves) at parameters $\Delta=1.0$, $Q=1.5$,
$D=1$}
\end{figure}

\end{document}